# Demographic Bias of Expert-Level Vision-Language Foundation Models in Medical Imaging

Yuzhe Yang[1], Yujia Liu[2], Xin Liu[3], Avanti Gulhane[4], Domenico Mastrodicasa[4,5], Wei Wu[4], Edward J. Wang[2], Dushyant W. Sahani[4], Shwetak N. Patel[3,6]

**Affiliations:**

[1] Department of Electrical Engineering and Computer Science, Massachusetts Institute of Technology; Cambridge, MA, USA.
[2] Department of Electrical and Computer Engineering, University of California, San Diego; La Jolla, CA, USA.
[3] Paul G. Allen School of Computer Science & Engineering, University of Washington; Seattle, WA, USA.
[4] Department of Radiology, University of Washington School of Medicine; Seattle, WA, USA.
[5] OncoRad/Tumor Imaging Metrics Core (TIMC), Department of Radiology, University of Washington; Seattle, WA, USA.
[6] Department of Electrical and Computer Engineering, University of Washington; Seattle, WA, USA.

## Abstract

Advances in artificial intelligence (AI) have achieved expert-level performance in medical imaging applications. Notably, self-supervised vision-language foundation models can detect a broad spectrum of pathologies without relying on explicit training annotations. However, it is crucial to ensure that these AI models do not mirror or amplify human biases, thereby disadvantaging historically marginalized groups such as females or Black patients. The manifestation of such biases could systematically delay essential medical care for certain patient subgroups. In this study, we investigate the algorithmic fairness of state-of-the-art vision-language foundation models in chest X-ray diagnosis across five globally-sourced datasets. Our findings reveal that compared to board-certified radiologists, these foundation models consistently underdiagnose marginalized groups, with even higher rates seen in intersectional subgroups, such as Black female patients. Such demographic biases present over a wide range of pathologies and demographic attributes. Further analysis of the model embedding uncovers its significant encoding of demographic information. Deploying AI systems with these biases in medical imaging can intensify pre-existing care disparities, posing potential challenges to equitable healthcare access and raising ethical questions about their clinical application.

# Main

Artificial intelligence (AI) has increasingly been deployed in real-world clinical settings, especially for medical imaging[1–4]. The latest developments include vision-language foundation models that operate on a self-supervised learning paradigm[5,6], eliminating the need for explicit pathology annotations while maintaining human-level diagnostic accuracy across various modalities and disease conditions[5,7–9]. Notably in radiology, by simultaneously using image and text inputs and leveraging the information naturally present in clinical reports associated with radiology images, foundation models identify pathologies without dependence on specific annotations, achieving performance that matches the expertise of radiologists and, in some cases, surpasses the expected diagnostic benchmarks[10,11].

Despite the plausible performance in diagnosing unseen pathologies[10], the foundation model could amplify existing biases in the data, causing diagnosis disparities across protected subpopulations and leading to unequal predictive outcomes for specific demographics[12–14] (e.g., discrepancies in diagnosis rates between Black and White patients). Existing literature has revealed that chest X-ray classifiers trained to predict the presence of disease systematically underdiagnosed Black patients[12,14], potentially leading to incorrect triage decisions and delayed medical treatment. Although algorithmic biases have been studied in the supervised setting[14–16] (e.g., models trained for specific diseases like "No Finding") or image-only foundation model[17] (e.g., pretrained solely on chest X-rays and fine-tuned on labeled data), little attention has been paid to *vision-language foundation models*. These models, notably free from explicit supervision through multimodal training and zero-shot inference, theoretically have reduced potential to inherit human labeling biases. However, to ensure the responsible and fair deployment, it is essential to investigate potential biases these models may possess, understand the sources and measure their outcomes, and where possible, initiate corrective actions[18].

We present a systematic study to measure and understand biases in vision-language foundation models. Using chest X-rays as a driving example, we mainly utilize CheXzero[10], a state-of-the-art self-supervised foundation model in medical imaging, to assess bias and fairness across a broad spectrum of pathologies with demographic subpopulations present in the testing data. We also test another vision-language foundation model[11] and show similar findings (Extended Data Fig. 1). Our analysis incorporates five diverse, globally-sourced radiology datasets: MIMIC[19], CheXpert[20], NIH[21], PadChest[22], and VinDr[23]. We evaluate fairness within both individual and intersectional subpopulations spanning demographic attributes including race, sex, and age[12,14]. We further compare fairness outcomes of the model with board-certified radiologists, uncovering that the foundation model demonstrates more substantial fairness discrepancies compared to human experts (Fig. 1). Further investigation in direct assessment of demographic attributes from chest X-rays shows that the model exhibits enhanced capacity to predict sensitive demographic information (e.g., age, race) compared to radiologists. Our international evaluation highlights pronounced biases within foundation models contrasted with evaluations by board-certified radiologists, shedding light on the origins of these biases and potential methodologies for bias auditing and correction before real deployments.

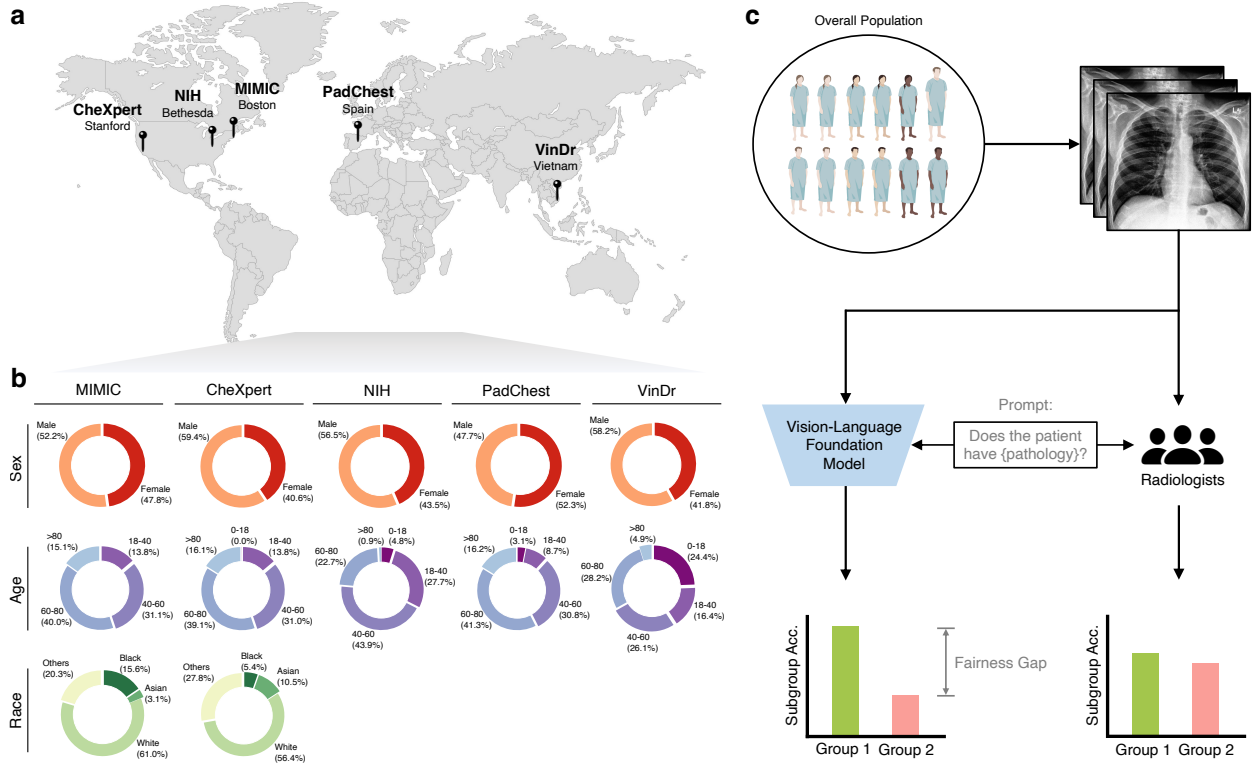

**Figure 1. The model evaluation pipeline. a,** We use internationally-sourced chest X-rays datasets for model evaluation, including MIMIC (Boston, MA), CheXpert (Stanford, CA), NIH (Bethesda, MD), PadChest (Spain), and VinDr (Vietnam). **b,** Distribution of demographics attributes (i.e., sex, age, race) of each dataset. For each attribute, we select common subgroups based on literature definition (sex: "male", "female"; age: "0-18", "18-40", "40-60", "60-80", ">80"; race: "Asian", "Black", "White", "Others"). Each dataset encompasses different proportions of subgroups, reflecting the diverse distributions in real-world clinical settings. **c,** For fairness evaluation, we processed radiographs through foundation models, accompanied by specific text prompts (e.g., "Does the patient have {pathology}?"). The evaluations are conducted across a wide range of different pathologies. Concurrently, board-certified radiologists independently reviewed identical subsets of the data, providing diagnoses that served as human fairness evaluations and comparisons (Fig. 2). In addition, we perform evaluations to assess the prediction of demographic attributes (e.g., sex, age, race) by both the model and three board-certified radiologists, following the same pipeline with modified prompts (Fig. 5).

# Results

## Datasets, Model, and Evaluation Protocols

We collect five public chest X-ray datasets from diverse global sources. These datasets, as detailed in Table 1, encompass MIMIC[19] (357,167 images from 61,927 patients), CheXpert[20] (223,458 images from 64,925 patients), and NIH[21] (112,120 images from 30,805 patients) from the United States, PadChest[22] (160,736 images from 67,590 patients) from Spain, and VinDr[23] (5,323 images from 5,323 patients) from Vietnam. The datasets provide chest X-ray images along with pathology labels and demographic data derived from the respective patients. Both MIMIC and CheXpert present demographic information including sex, age, and ethnicity. The remaining datasets (i.e., NIH, PadChest, VinDr) present demographic details regarding sex and age, with no information available on the race of the patients.

**Table 1. Characteristics of the datasets used in this study.**

| | | MIMIC | CheXpert | NIH | PadChest | VinDr |
|---|---|---|---|---|---|---|
| Location | | Boston, MA | Stanford, CA | Bethesda, MD | Alicante, Spain | Hanoi, Vietnam |
| # Images | | 357,167 | 223,458 | 112,120 | 160,736 | 5,323 |
| # Patients | | 61,927 | 64,925 | 30,805 | 67,590 | 5,323 |
| # Frontal | | 230,406 | 191,014 | 112,120 | 111,187 | 5,323 |
| # Lateral | | 126,761 | 32,444 | 0 | 49,549 | 0 |
| # Pathologies | | 14 | 14 | 15 | 174 | 27 |
| Race | Asian | 11,121 | 23,384 | | | |
| | Black | 55,611 | 11,999 | | | |
| | White | 218,037 | 125,990 | | | |
| | Other | 72,398 | 62,085 | | | |
| Sex | Female | 170,698 | 90,833 | 48,780 | 79,880 | 2,227 |
| | Male | 186,469 | 132,625 | 63,340 | 80,856 | 3,096 |
| Age | 0-18 | 0 | 103 | 5,402 | 5,529 | 1,298 |
| | 18-40 | 49,353 | 30,808 | 31,037 | 14,033 | 874 |
| | 40-60 | 111,055 | 69,245 | 49,243 | 41,406 | 1,390 |
| | 60-80 | 142,824 | 87,378 | 25,419 | 62,206 | 1,501 |
| | 80-100 | 53,935 | 35,924 | 1,019 | 37,562 | 260 |

We utilize a state-of-the-art self-supervised foundation model in medical imaging, CheXzero[10], as a driving example to study fairness of foundation models. The details about model architecture and training are in the Methods section. Note that the model was trained in a self-supervised way without using any pathology labels or annotations. We also tested on another vision-language foundation model[11] and observed similar findings (Extended Data Fig. 1). We evaluated the model on internationally-sourced chest X-rays datasets. In particular, CheXpert, PadChest, and VinDr contain gold-standard ground truth radiologist labels. Among these datasets, CheXpert test set (666 samples) and VinDr (5,323 samples) provide external annotations from three board-certified radiologists, which were used to benchmark the performance and fairness of radiologists' assessments compared to the model.

To assess the model prediction fairness, we focus on three demographic attributes: sex, age, and race, and dissect the performance of the model within different subpopulations, such as female or Black patients, and the intersectional groups like Black female patients. We follow the literature to examine the class-conditioned error rate that is likely to lead to worse patient outcomes for a screening model[12,14]. For all potential pathology labels, a false negative indicates falsely predicting someone to be healthy when they are ill, which could lead to delays in treatment[14] (i.e., an underdiagnosis). Therefore, we evaluate the differences in False Negative Rate (FNR) between demographic subpopulations. For "No Finding", we evaluate the False Positive Rate (FPR) for the same reason. Equality in these metrics can be viewed as instances of equal opportunity between subgroups[24]. We then denote the differences in FNR/FPR for two selected subgroups (e.g., Black and White patients) as the ***underdiagnosis disparity***.

**Model Exhibits Larger Fairness Disparities Compared to Radiologists**

We assess the model's underdiagnosis disparity across datasets and demographic populations. Since external radiologist annotations are available in certain datasets (i.e., CheXpert and VinDr), we directly compared the overall performance as well as the performance for subpopulations between the model and radiologists. Fig. 2 presents the diagnostic performance and fairness of the vision-language foundation model in contrast to that of board-certified radiologists on the CheXpert dataset (n=666). First, Figs. 2a, 2c, and 2e show the comparison of the receiver operating characteristic (ROC) curves of the model to the operating points of radiologists for three different pathologies. Notably, the model exhibits comparable or better diagnostic performance compared to radiologists ("Enlarged Cardiomediastinum": AUC=0.917, 95% CI [0.905, 0.928]; "Pleural Effusion": AUC=0.938, 95% CI [0.922, 0.950]; "Lung Opacity": AUC=0.919, 95% CI [0.904, 0.933]).

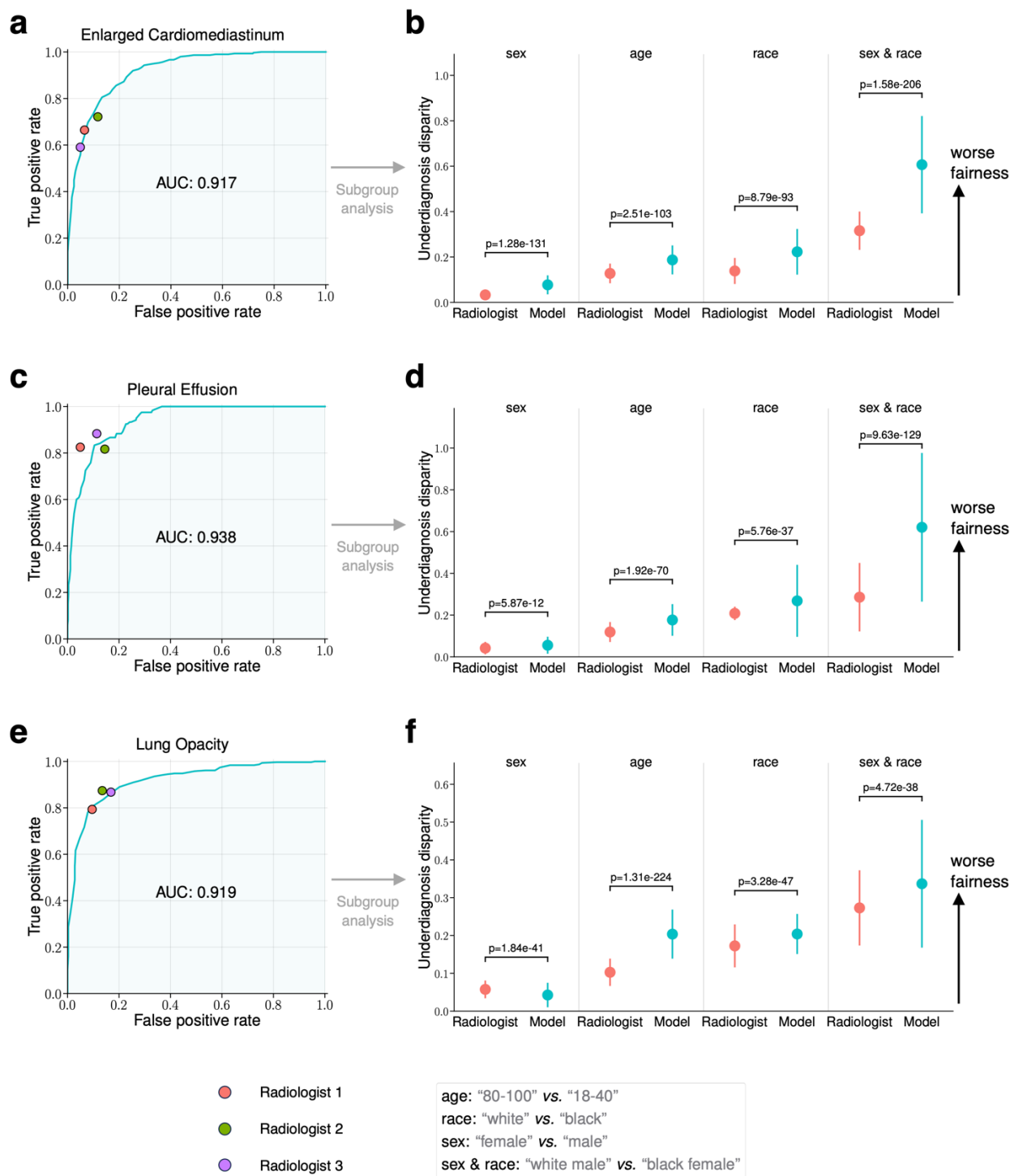

**Figure 2. Comparisons of diagnosis AUROC and underdiagnosis disparity for the vision-language foundation model and board-certified radiologists. a, c, e,** Comparison of the ROC curve of the vision-language foundation model to benchmark radiologists against the test-set ground truth on the CheXpert dataset (n=666). The model outperforms the radiologists when the ROC curve lies above the radiologists' operating points. The model has an AUC of 0.917 (95% CI [0.905, 0.928]) for "Enlarged Cardiomediastinum", an AUC of 0.938 (95% CI [0.922, 0.950]) for "Pleural Effusion", an AUC of 0.919 (95% CI [0.904, 0.933]) for "Lung Opacity". **b, d, f,** Comparison of the underdiagnosis disparity of the vision-language foundation model against three

board-certified radiologists on the CheXpert test set (n=666). We average the assessments from different radiologists as the evaluation of human biases. The model exhibits significantly higher underdiagnosis bias than that of radiologists on all three pathologies. Error bars indicate 95% confidence intervals estimated using non-parametric bootstrap sampling (n=1,000) are shown. More results can be found in the Extended Data Fig. 2.

In the meantime, we further assess the underdiagnosis disparity between subgroups, which measures the disparity of FNR between two selected subgroups in each category ("female" vs "male" in sex, "80-100" vs "18-40" in age, "white" vs "black" in race, and "white male" vs "black female" in the intersectional group of sex and race). We average the assessments from different radiologists as the evaluation of human biases. When computing FNR for the model, we use the optimal threshold computed on the validation set that maximizes the Youden's J statistic[25]. Figs 2b, 2d, and 2f show that the model exhibits much larger fairness gaps compared to radiologists, especially for intersectional subgroups. For instance, the model exhibits significantly higher underdiagnosis rate for "Enlarged Cardiomediastinum" in sex (p=1.28e-131, one-tailed Wilcoxon rank-sum test; same test for following attributes), age (p=2.51e-103), race (p=8.79e-93), and the intersectional of sex and race (p=1.58e-206). More results can be found in the Extended Data Fig. 2, including the analysis of other pathologies in CheXpert, and on another dataset from a different site (VinDr). Overall, the model exhibits expert-level pathology detection accuracy, but shows consistently higher underdiagnosis bias compared to radiologists.

**Diagnosis Bias in Marginalized Populations and Intersectional Groups**

We further evaluate the diagnosis bias of the model on MIMIC, the largest and the most diverse chest X-ray dataset in our study. We focus on the "No Finding" label, and show both underdiagnosis and overdiagnosis bias of the model on individual and intersectional subpopulations (Fig. 3). FPR is used for assessing underdiagnosis, whereas FNR is used for overdiagnosis. Fig. 3a shows significant fairness gaps between patient subpopulations in each category, especially between the age subgroups ">80" (n=53,935) and "18-40" (n=49,353). Moreover, larger gaps of the underdiagnosis rate between the intersectional subgroups can be observed in Fig. 3b. For instance, around 20% FPR discrepancies exist between female patients aged above 80 (n=29,209) and those in their 18-40 (n=25,350). Similar observations hold for overdiagnosis (Figs. 3c, 3d), the gaps become more significant between intersectional subgroups. The FPR (Fig. 3a) and FNR (Fig. 3c) for "No Finding" shows an inverse relationship across different marginalized subgroups in the CXR dataset. Such an inverse relationship also exists for intersectional subgroups (Figs. 3b, 3d), and is consistent across other datasets.

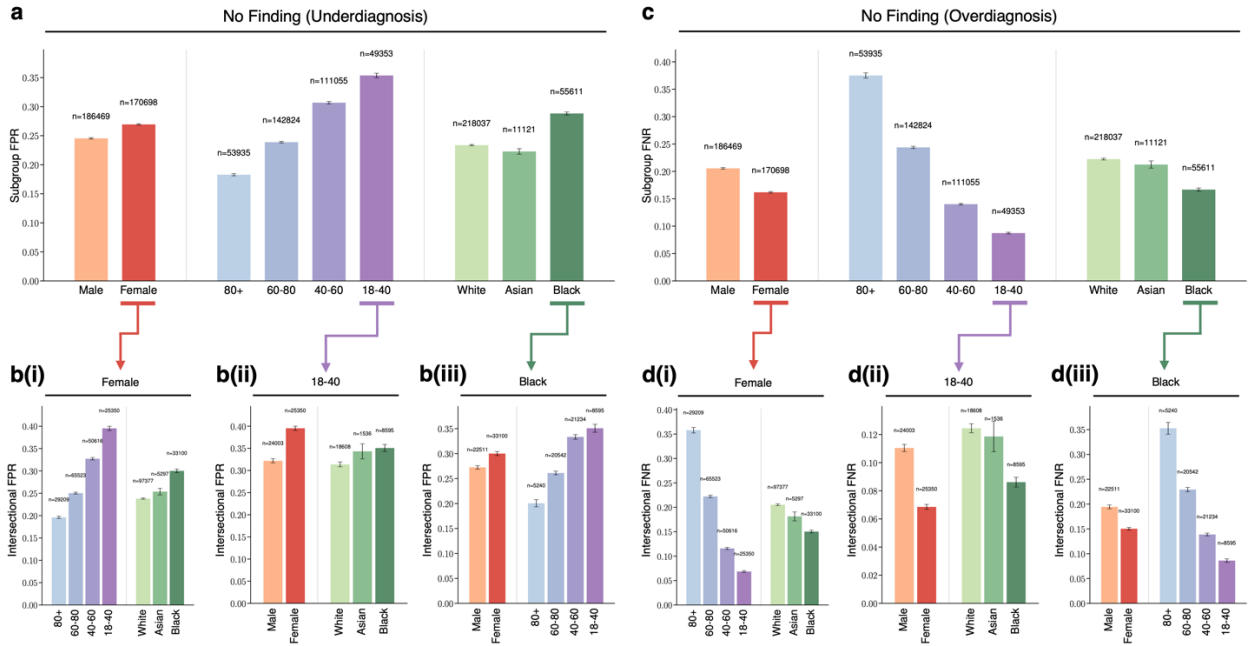

**Figure 3. Analysis of underdiagnosis and overdiagnosis across subgroups of sex, age, race, and intersectional groups in the MIMIC dataset. a,** The underdiagnosis rate, as measured by the no finding FPR, in the indicated patient subpopulations. **b,** Intersectional underdiagnosis rates for female patients (**b(i)**), patients aged 18–40 years (**b(ii)**), and Black patients (**b(iii)**). **c, d,** The overdiagnosis rate, as measured by the no finding FNR in the same patient subpopulations as in **a** and **b**. Error bars indicate 95% confidence intervals estimated using non-parametric bootstrap sampling (n=1,000) are shown. More results can be found in the Extended Data Fig. 3.

We observe that female patients, patients aged between 18 and 40 years, and black patients have higher rates of algorithmic underdiagnosis, indicating that these subgroups are most likely being falsely diagnosed as healthy by the model and failing to receive appropriate clinical treatments. Further investigations on intersectional subpopulations reveal that the underdiagnosis rates increased significantly for specific groups of patients, such as black female patients. We show in Extended Data Fig. 3 that the observations hold across different pathologies such as "Lung Opacity" or "Pneumonia".

**Demographic Bias in Unseen Radiographic Findings**

We extended our analysis to investigate the demographic biases using a much larger and diverse set of pathology labels. We tested the foundation model on the PadChest dataset collected from a different country with 174 radiographic findings and 19 differential diagnoses[22]. We filtered out 48 radiographic findings where n>100 and the model achieved an AUC of at least 0.7 in the PadChest test set (n=39,053) to further assess the demographic fairness of the model on unseen radiographic findings[10]. Fig. 4 reveals distinct disparities in both sex ("female" vs "male" subgroup) and age (">80" vs "18-40" subgroup) among those radiographic findings. The maximum underdiagnosis disparity (i.e., "Multiple nodules", n=102) between female and male patients is 24.1% (95% CI [22.5%, 26.0%]), whereas 31 out of 48 findings exhibit a fairness gap larger than 5% (Fig. 4a). The discrepancies become even more significant for age, with a 100% fairness gap for "Tracheostomy tube" (n=163) between "18-40" and ">80" subgroups, and 45 out of 48 findings exhibit a fairness gap larger than 20% (Fig. 4b).

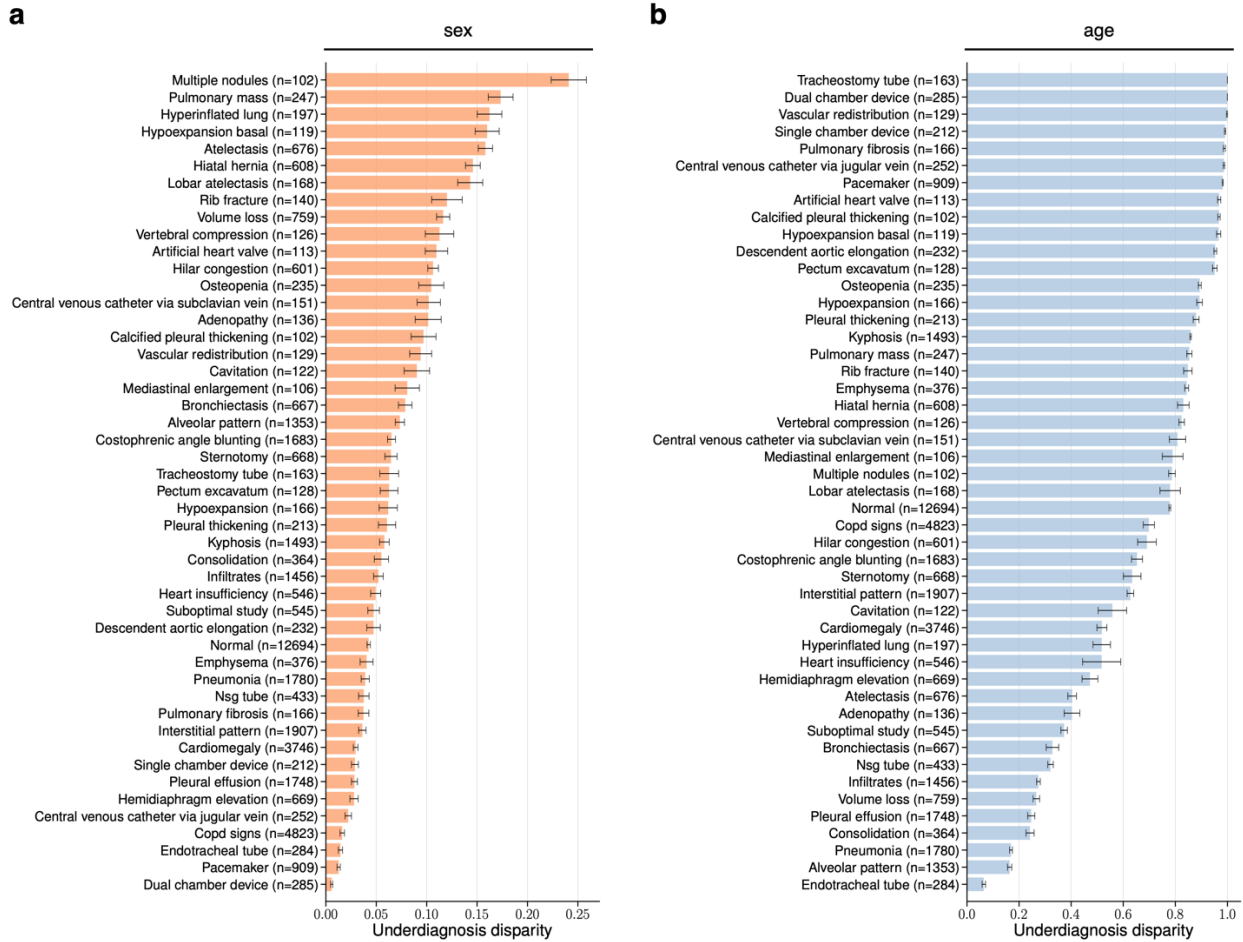

**Figure 4. Demographic fairness on unseen radiographic findings in the PadChest dataset.** Average underdiagnosis disparity and 95% CI are shown for each radiographic finding (n>100) labeled as high importance by an expert radiologist. **a,** Underdiagnosis disparity for sex (between group "female" and "male". **b,** Underdiagnosis disparity for age (between group "18-40" and ">80". We externally validated the model's fairness when testing on different data distributions by evaluating model performance on the human-annotated subset of the PadChest dataset (n= 39,053). No labeled samples were seen during training for any of the radiographic findings in this dataset.

**Foundation Model Encodes Demographic Information Beyond Human Levels**

With consistent demographic bias across international evaluation, we aim to further dissect and explain the performance of the model. Inspired by recent works on algorithmic encoding of demographic information by deep learning models[26–28], we investigated whether the model encodes demographic information by examining the predictability of sensitive attributes by both the self-supervised foundation model and board-certified radiologists. We selected 480 chest X-ray samples from the MIMIC dataset, ensuring an equal number of samples across all subgroups in three key attributes: sex, age, and race (details in Methods). Instead of focusing on pathology prediction, we assessed how much the model encodes demographic information by training a linear attribute prediction head using logistic regression on top of the penultimate layer of the model, with the model weights frozen. In the meantime, we recruited three board-certified radiologists to label the demographic attributes (sex, age, and race) for the same set of patients based solely on their chest X-rays (details in Methods).

Interestingly, the foundation model, although trained in a self-supervised manner without explicit information regarding the demographic attributes, demonstrated substantial and consistent encoding of demographic information across all tested attributes and subgroups (Fig. 5). Specifically, the predictive AUCs for sex (Fig. 5a, "female" AUC=0.92, 95% CI [0.91, 0.93]), age (Fig. 5b, "18-40" AUC=0.94, 95% CI [0.93, 0.94]), race (Fig. 5c, "black" AUC=0.78, 95% CI [0.77, 0.78]), and the intersectional subgroups (Fig. 5d, "black female" AUC=0.83, 95% CI [0.82, 0.83]) are significantly higher than random chance (i.e., 0.5). This strong algorithmic encoding of demographic attributes could be explainable for the observed underdiagnosis bias across patient subpopulations (details in Discussion).

Interestingly however, the performance of three radiologists to predict these attributes falls behind. They achieve relatively high AUC scores in sex prediction (Fig. 5a), but much lower in age prediction (Fig. 5b). When it comes to race, the prediction is marginally better than random guess (Fig. 5c). Similar performance pattern is observed in the intersectional group of sex and race prediction (Fig. 5d), suggesting that radiologists cannot directly read attributes like age or race from radiographs. Complete results on other datasets are in Extended Data Fig. 4. We provide further analysis and initial methods to intervene the model fairness in Extended Data Figs. 5 and 6.

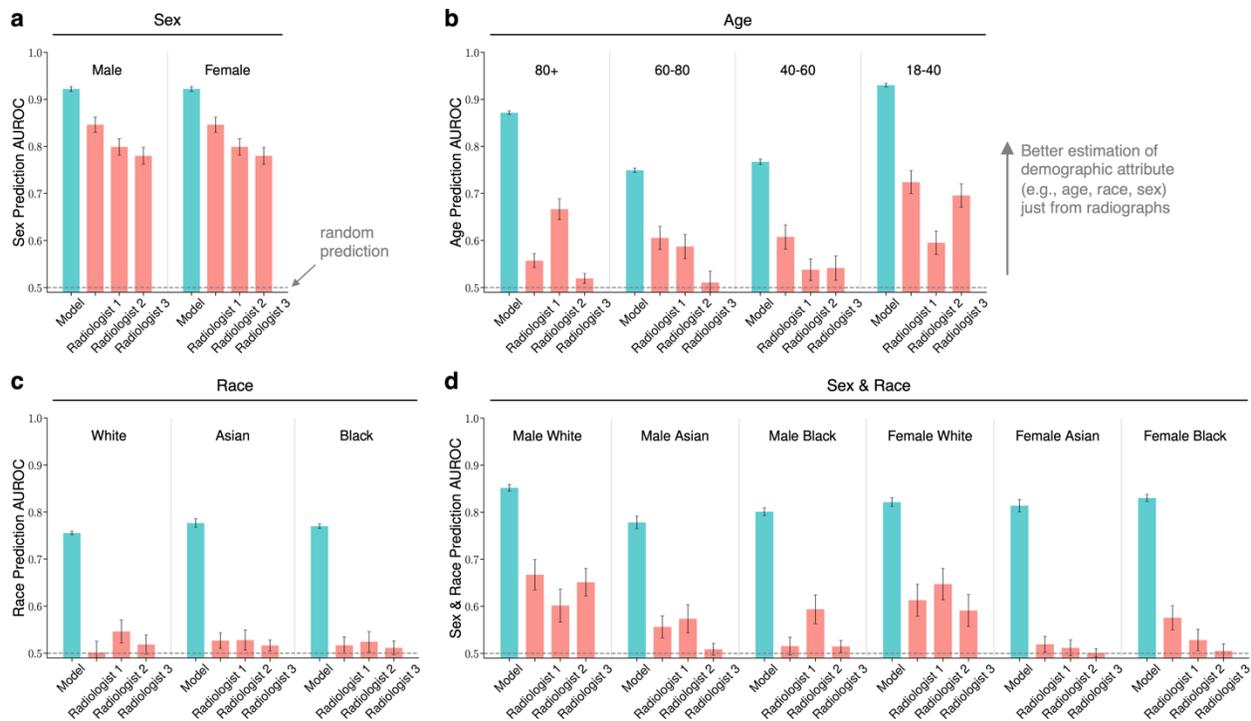

**Figure 5. Comparisons of prediction AUROC for sensitive demographic attributes between the foundation model and three board-certified radiologists. a, b, c, d,** Prediction AUROC of subgroups within different sensitive attributes including sex (**a**), age (**b**), race (**c**), and the intersectional groups of sex and race (**d**), on a subset of MIMIC (n=480). We selected out a balanced subset of MIMIC w.r.t. all attributes (i.e., balanced across age, sex, and race), and asked three board-certified radiologists to infer the attributes from just the chest X-rays. To assess the model prediction of sensitive attributes, we train a linear attribute prediction head using logistic regression on top of the penultimate layer of the model, with the model weights frozen. Error bars indicate 95% confidence intervals estimated using non-parametric bootstrap sampling (n=1,000) are shown. Complete results of model predictions for other datasets are in Extended Data Fig. 4.

## Discussion

We have dissected the performance of the state-of-the-art foundation model and shown consistent underdiagnosis in five internationally-sourced public datasets in the chest X-ray domain. Importantly, we were able to compare the results with board-certified radiologists to ground the findings. The results reveal consistently larger fairness disparities of the model compared to radiologists (Fig. 2), and that the model exhibits systematic underdiagnosis biases in marginalized subpopulations, such as female, younger, Black patients, as well as intersectional subgroups like Black female patients (Fig. 3). The demographic biases of the foundation model also persist across a wide range of unseen pathologies (Fig. 4). Further analyses show that the model encodes substantial demographic information (e.g., race), and that is significantly higher than human radiologists (Fig. 5).

Our results have multiple implications. First, the fairness-accuracy trade-off in AI models can raise complex ethical considerations[29,30]. The latest advancements in medical vision-language foundation models hold the promise of a single model diagnosing countless pathologies with expert-level accuracy. Yet, our analysis shows that they exhibit significant fairness gaps over marginalized groups. This disparity is significantly larger than that by radiologists across various diagnostic tasks and patient subpopulations (Fig. 2). Incorrectly underdiagnosing specific subgroups (e.g., black female patients) more frequently than others not only places these individuals at a disadvantage but also raises serious ethical concerns when deploying the model in a clinical pipeline[31,32]. The results have implications on the regulation of these new medical technologies[33], especially under the recent White House Executive Order on the safe, secure, and trustworthy development and use of artificial intelligence[34].

Second, our study shows that the model encodes demographic information far more profoundly than human capacity (Fig. 5, Extended Data Fig. 4). This suggests that inherent encoding of the sensitive data might drive the underdiagnosis biases. Notably, even though the model is trained in a self-supervised manner without explicit attribute information, it still manages to embed this information. Recent studies explore if deep models use demographics as "shortcuts", disadvantaging specific groups[35,36]. These call for a deeper understanding of how these powerful models process and utilize sensitive information, and whether that is aligned with clinical validations by radiologists. While demographics can refine differential diagnoses and are associated with patient outcomes in certain cases, they may not be a direct causal factor in most diseases[32,37] (e.g., Pneumothorax, Pneumonia, Fracture, etc.). Whether demographic variables should be encoded as proxies for causal factors is a decision that should align with its actual clinical use[18,31,38].

Third, our results reveal that the AI model effectively encodes demographic information from radiographs more accurately than radiologists. A recent study demonstrates that AI can measure and interpret biological age, predict age-related outcomes, and convert these predictions into an estimated biological age[39]. This capability extends beyond the typical scope of radiologist assessments, which generally do not include evaluating age, sex, or gender, as these details are usually obtained from electronic medical records. However, the ability of the AI model to discern these demographics more precisely suggests a potential for uncovering new and potentially clinically relevant features that might not be immediately apparent to human practitioners, paving the way for enhanced human-AI collaboration[40–42]. Delving into the semantic and agnostic features harnessed by AI could enrich human comprehension and foster medical education. Such findings emphasize the need for further exploration into human-AI interactions for a holistic understanding and better collaborations. On the other hand, in scenarios where clinical decisions are influenced by AI model suggestions, any undetected bias within the model could lead to unintended and potentially harmful consequences[29,43]. This underscores the need for careful and continuous evaluation of AI biases to progressively diminish their influence in healthcare, and ensure more accurate and equitable diagnoses.

Fourth, our human study reveals that radiologists too can manifest biases when evaluating over diverse subgroups. These inherent biases raise pressing concerns about potential corrective actions and the mechanism for feedback. Existing literature points out that clinician bias significantly contributes to healthcare disparities across race and gender[44–46]. With the addition of biases from AI, the need for de-identifying demographics to counteract biases becomes vital. This is especially crucial as AI demonstrates the capability to uncover demographics even without such supervision. Concurrently, regulatory authorities should establish clear guidelines on how the prediction of demographics (among other things) should be managed by AI models, ensuring they adhere to privacy and ethical standards.

Our study also has some limitations. First, demographics associated with the datasets are mainly self-reported or physician-recorded, where inconsistencies during examinations can introduce label noise. Demographic labels can be influenced by numerous characteristics such as age, socioeconomic status, and levels of cultural assimilation. Mitigating such label noise remains challenging as traditional bias mitigation strategies may not provide effective corrections, leading to inherent biases embedded in data and models. Second, the datasets used in this study included a range of chest X-rays and projections, many of which did not adhere to a uniform standard of image quality. The extent to which this variability might have impacted our results remains unclear and highlights the need for future research to explore the influence of image quality on AI model performance. Third, we focused primarily on one specific vision-language foundation model implementation. While the model is considered state-of-the-art and serves as a robust starting point, the methodologies used in the study are generic and adaptable to other foundation models[7–10]. Lastly, while we involve human studies to assess the diagnostic performance and fairness of radiologists, they face constraints due to the limited number of participating radiologists. Expanding the scale of participants in future studies could improve the validity of the results.

In summary, we have uncovered pronounced and systematic demographic biases of the state-of-the-art visual-language foundation model, and performed human evaluations to compare them to radiologists. The results, validated across five large-scale, globally-sourced datasets, showed that even when AI achieves human-level performance, deploying these algorithms in real-world scenarios demands careful consideration of the ethical implications, especially concerning underrepresented subpopulations. Furthermore, our findings prompt new questions into how human-AI collaborations might evolve, harnessing insights beyond human comprehension to foster more effective human-AI interactions for enhanced patient care.

## Methods

**Datasets and Pre-Processing**

We provide additional information about the datasets used in this study. The datasets are summarized in Table 1. All five public datasets offer demographic attributes of sex and age for the associated patients. MIMIC and CheXpert additionally provide demographic information on race. To ensure the integrity of the datasets, we exclude the samples with incomplete demographic data from the dataset. Specifically, if any of the essential attributes: sex, age, or race (for MIMIC and CheXpert) is missing, the corresponding image is excluded from consideration in our study. Notably, the reported numbers in the paper regarding the datasets are post application of this exclusion criteria.

In total, we have 357,167 images from MIMIC (MIMIC-CXR), 223,458 images from CheXpert, 112,120 images from NIH (ChestX-ray14), 160,736 images from PadChest, and 5,323 images from VinDr (VinDr-CXR).

**MIMIC.** The MIMIC (MIMIC-CXR) dataset[19] contains 357,167 chest X-rays along with free-text radiology reports, obtained from Beth Israel Deaconess Medical Center (BIDMC) in Boston, MA. The chest radiographs are retained in Digital Imaging and Communications in Medicine (DICOM) format, and the radiology reports are extracted from BIDMC electronic health record (EHR) which include textual descriptions and interpretations of the findings in the X-ray images written by radiologists during routine care.

**CheXpert.** The CheXpert dataset[20] consists of 223,458 chest X-rays from 64,925 patients. It offers a diverse and comprehensive collection of chest radiographs that span a wide array of subpopulations. An automated rule-based labeler was developed to extract the 14 observations (radiographic findings) from the radiology reports. Annotations from board-certified radiologists are available for both the validation set (200 studies sampled randomly from the full dataset) and test set (500 studies randomly sampled from the 1,000 studies in the report evaluation set for the labeler). Three of the eight board-certified radiologists were chosen to benchmark the performance of radiologists[20].

**NIH.** The NIH dataset (ChestX-ray14)[21] is a medical imaging dataset that contains 112,120 frontal-view chest X-ray images. These images are sourced from 30,805 patients, with data collected over a large period ranging from 1992 and 2015. It provides 15 pathology labels extracted from the radiological reports through text mining techniques. As an expansion of ChestX-ray8, this dataset introduces six additional thoracic diseases: Edema, Emphysema, Fibrosis, Pleural Thickening, and Hernia. The chest X-ray images are resized to a resolution of 1024x1024 pixels from the original DICOM format.

**PadChest.** The PadChest dataset[22] comprises a substantial collection of 160,736 chest X-ray images obtained from 67,590 patients at San Juan Hospital (Spain) from 2009 to 2017. It offers six different radiographic projections, 174 findings, and 19 differential diagnoses. 39,053 chest X-ray images were manually annotated by trained physicians, and a recurrent neural network with attention mechanism trained on the manually labeled subset was used to label the remaining samples.

**VinDr.** The VinDr (VinDr-CXR) dataset[23] is a public dataset comprising 5,323 frontal chest X-ray images collected from 5,323 patients across two major hospitals in Vietnam: Hospital 108 (H108) and Hanoi Medical University Hospital (HMUH). The dataset provides labels for 27 findings (the label "Other diseases" is removed from the original dataset to avoid ambiguity). Notably, it is considered a high-quality dataset of annotated images in the research community as it provides radiologists-generated annotations for all images (in both training and test sets).

**Model Training, Evaluation, and Human Study Details**

We mainly employ the CheXzero model[10] as a driving example to study fairness of foundation models. The vision-language model was initialized from a Vision Transformer backbone ViT-B/32[47] and pre-trained weights from OpenAI's CLIP model which excels in tasks related to vision and language understanding[48]. The model was trained in a self-supervised manner on the MIMIC dataset with no pathology labels or annotations used, just by leveraging the radiographs with accompanying clinical texts[10]. In addition, we also tested another vision-language foundation model, KAD[11], which introduces knowledge graphs into visual-language pretraining.

We evaluated the model on our internationally-sourced chest X-ray datasets. In particular, approximately 45,000 chest X-ray images used in our evaluation come with gold standard annotations from radiologists across three datasets: CheXpert test set (666 chest X-rays with eight board-certified radiologist annotations for the presence of 14 different conditions), VinDr (5,323 images with annotations from a total of 17 experienced radiologists for 27 findings and diagnoses), and a subset of PadChest (39,053 images from the original dataset annotated by trained physicians). We also tested the model performance and fairness on MIMIC (357,167 images) and NIH (112,120 images) where the labels are generated from natural language processing techniques. Following the standard preprocessing practice[4,49], we resized the radiographs to 224x224 and normalized them using a sample mean and standard deviation of the dataset for model evaluation.

We recruited three board-certified radiologists from the Department of Radiology at the University of Washington, School of Medicine to evaluate the predictability of demographic attributes from radiographs only. We utilized an online labeling tool[50] for the radiologists to create attribute labels based on the 480 pre-selected chest X-ray images from MIMIC. All three attribute labels are required for each image, meaning that radiologists are required to choose one label for each of the attributes: sex (female, male), age (0-18, 18-40, 40-60, 60-80, >80), and ethnicity (asian, black, white, others). Importantly, each radiologist completed this study independently and was provided with no additional information beyond the chest X-ray images themselves. The distribution of the 480 data samples across the three attributes was not disclosed to the radiologists until after they had completed the task, ensuring an unbiased evaluation process.

**Evaluation Methods**

To evaluate the performance of the foundation model on pathology classification, we use the following metrics: true positive rates (TPR), true negative rates (TNR), receiver operating characteristic (ROC) curves, and the area under the ROC curve (AUC). To evaluate the underdiagnosis disparity given one demographic attribute, we use the difference in TNR (or TPR) between two specific subpopulations (e.g., Black and White patients). To evaluate and assess the learned features in the penultimate layer of the model, we employ Principal Component Analysis (PCA)[51] to project the embeddings into a two-dimensional space for visualization.

TPR and TNR are calculated as (TP: True Positive; FN: False Negative; TN: True Negative; FP: False Positive):

$$TPR = \frac{TP}{TP + FN}$$
$$TNR = \frac{TN}{TN + FP}$$

We plotted ROC curves that demonstrate the trade-off between TPR and TNR as the classification thresholds are varied. When reporting the TPR and TNR, we used the optimal threshold computed on the validation set that maximizes the Youden's J statistic[25]. We followed standard non-parametric bootstrap sampling (n=1,000) to calculate the 95% confidence interval[52]. We also reported AUC, which is the area under the corresponding ROC curves showing an aggregate measure of detection performance.

**Assessing the Demographic Fairness of the Model**

To measure the fairness of the foundation model, we evaluate the metrics described above for each demographic subpopulation (defined over demographic attributes including sex, age, and race), and the differences in metric outcomes across these groups. The principle of equal TPR and TNR across different demographic subgroups is known as equal odds[24], a concept well-established in algorithmic fairness[53,54]. Given that the models we investigate in this work will likely be used as screening or triage tools, it is crucial to recognize that the cost of a False Positive (FP) may vary considerably from that of a False Negative (FN). Specifically, for a specific pathology, FNs (corresponding to underdiagnosis[14]) would be more costly than FPs, and so we focus on the FNR (or TPR) for this task. For the task of "No Finding" prediction, we focus on the FPR (or TNR) for the same reason. Equality in one of the class conditioned error rates is an instance of equal opportunity[24]. Consequently, we calculate underdiagnosis disparity as the difference in TNR (or TPR) between two selected subgroups.

## Additional Evaluation Results

**Assessing the encoding of attributes by text prompts.** We assessed the algorithmic encoding of demographic attributes in the foundation model through a logistic regression layer on the top of the model embedding (Fig. 5). Since the foundation model also supports textual prompts as input, we assess the encoding again by directly using textual prompts (Extended Data Fig. 5). Specifically, we utilized prompts containing demographic information (e.g., "The patient's gender is male.") to assess the attribute prediction accuracy. Across different datasets, the resulting prediction AUC is lower than using logistic regression, but still significantly higher than random chance over most of the subgroups.

**Model fairness intervention.** We conducted experiments to explore fairness intervention of the foundation model by incorporating demographic details into the input prompt (Extended Data Fig. 6). We proposed to intervene the model prediction over subgroups by including demographic information in the input texts (e.g., "Does this *female* patient have Pneumonia?"). Extended Data Fig. 6 shows complex outcomes: After such intervention, the model displays reduced demographic biases for certain conditions like "Lung Opacity" and "No Finding" (Extended Data Figs. 6a, 6c), but not for others like "Pneumonia" (Extended Data Fig. 6b). The results indicate that it is possible to improve the demographic fairness of the model while maintaining the overall performance, but deeper analyses are needed for more principled methods.

## Statistical Analysis

**Underdiagnosis Disparity.** One-tailed Wilcoxon rank-sum test ($\alpha=0.05$) was used to assess the underdiagnosis disparity between the foundation model and radiologists.

**AUROC.** We collect AUROC results for pathology prediction across five datasets using the foundation model's predictions. We also present the AUROC results for pathology prediction from external board-certified radiologists on the CheXpert test set (n=666) and VinDr test set (n=5,323). We collect AUROC results for attribute prediction (e.g., race) across five datasets using the foundation model's predictions with textual input changed to attribute predictions. We also present the AUROC results for attribute prediction from three board-certified radiologists on the subset of MIMIC (n=480). 95% CI for the true AUCs were estimated using non-parametric bootstrap sampling (n=1,000).

**TPR and TNR.** The mean metric is reported along with 95% confidence intervals, which are estimated using non-parametric bootstrap sampling (n=1,000).

**Confidence intervals.** We use the non-parametric bootstrap sampling to generate confidence intervals: random samples of size n (equal to the size of the original dataset) are repeatedly sampled 1,000 times from the original dataset with replacement. We then estimate the AUC, subgroup TNR

(or TPR) and underdiagnosis disparity (fairness gaps) metrics using each bootstrap sample ($\alpha=0.05$).

All statistical analysis was performed with Python version 3.9 (Python Software Foundation).

**Data Availability**

All datasets used in this study are publicly available. The MIMIC and VinDr datasets are available from PhysioNet after the completion of a data use agreement and a credentialing procedure. The CheXpert dataset, along with associated race labels, is available from the Stanford AIMI website. The ChestX-ray14 (NIH) dataset is available to download from the National Institute of Health Clinical Center. The PadChest dataset can be downloaded from the Medical Imaging Databank of the Valencia Region.

**Code Availability**

Code that supports the findings of this study is publicly available with an open-source license at https://github.com/YyzHarry/vlm-fairness.

**Acknowledgements**

Figure 1 was created with biorender.com.

# Extended Data

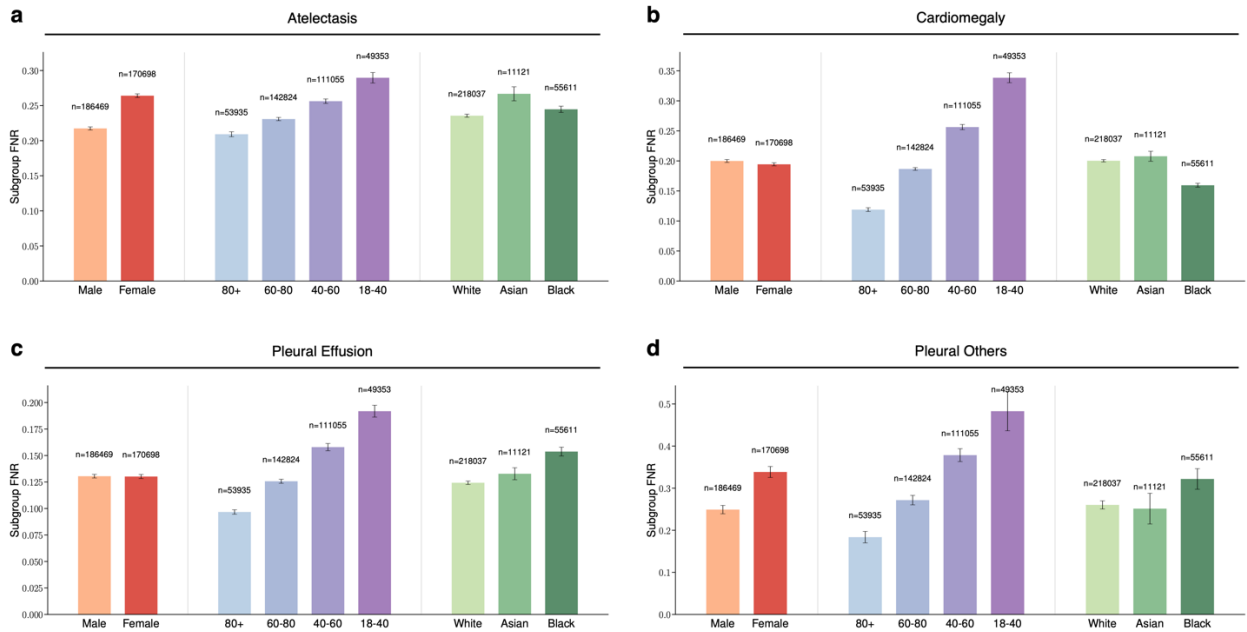

**Extended Data Figure 1. Underdiagnosis disparities on different pathologies of another vision-language foundation model, KAD[11], across subgroups of sex, age, and race in the MIMIC dataset. a, b, c, d,** The underdiagnosis rate for "Atelectasis", "Cardiomegaly", "Pleural Effusion", and "Pleural Others" in the indicated patient subpopulations. Error bars indicate 95% confidence intervals estimated using non-parametric bootstrap sampling (n=1,000) are shown.

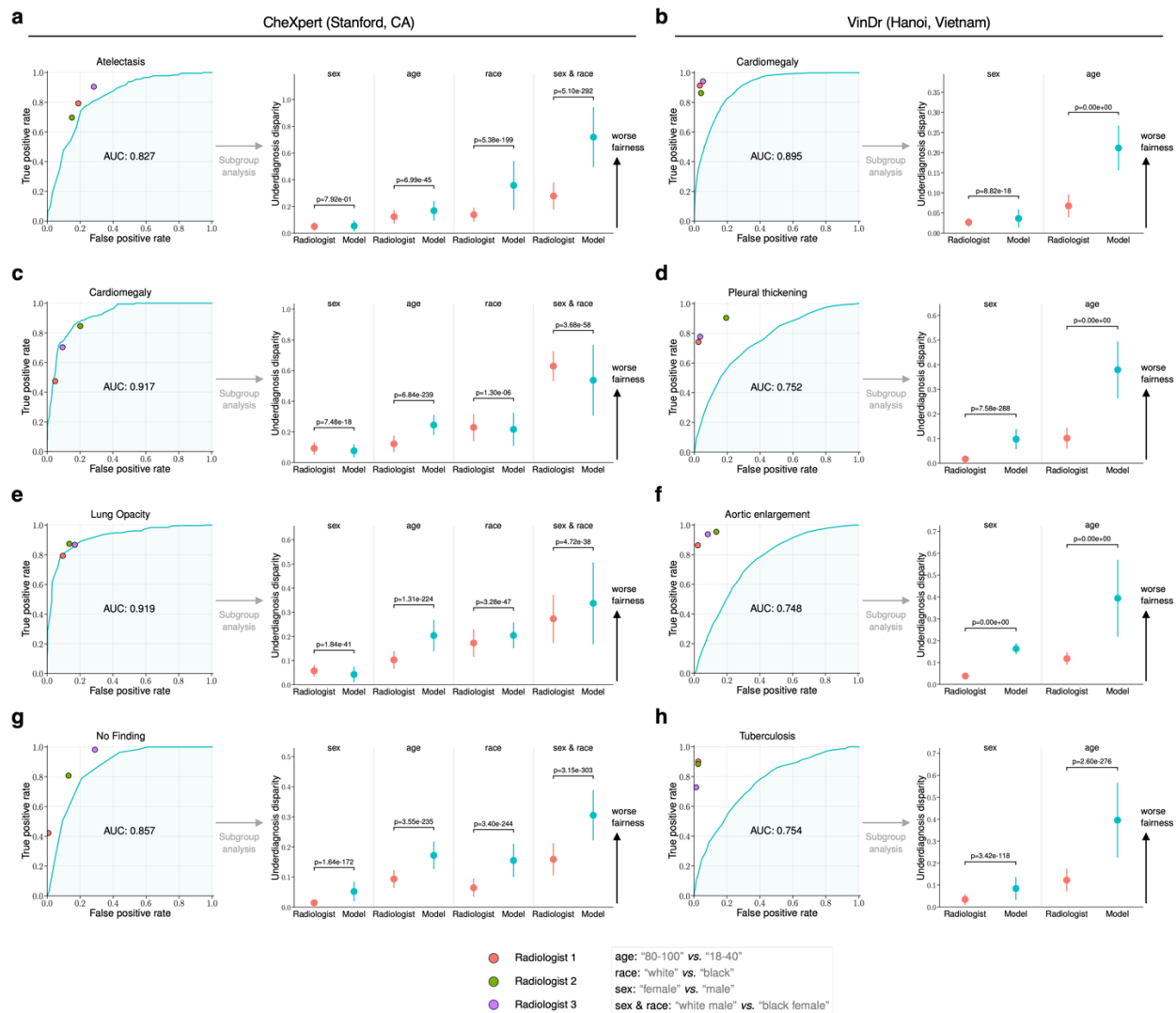

**Extended Data Figure 2. Comparisons of diagnosis AUROC and underdiagnosis disparity for the vision-language foundation model and board-certified radiologists on different datasets. a, c, e, g,** Comparison of the ROC curve (left) and the underdiagnosis disparity (right) of the model to benchmark radiologists against the test-set ground truth on the CheXpert dataset (n=666). **b, d, f, h,** The same comparisons performed on another dataset from a different country, VinDr (n=5,323). We average the assessments from different radiologists as the evaluation of human biases. The model exhibits significantly higher underdiagnosis bias than that of radiologists on all three pathologies. Error bars indicate 95% confidence intervals estimated using non-parametric bootstrap sampling (n=1,000) are shown.

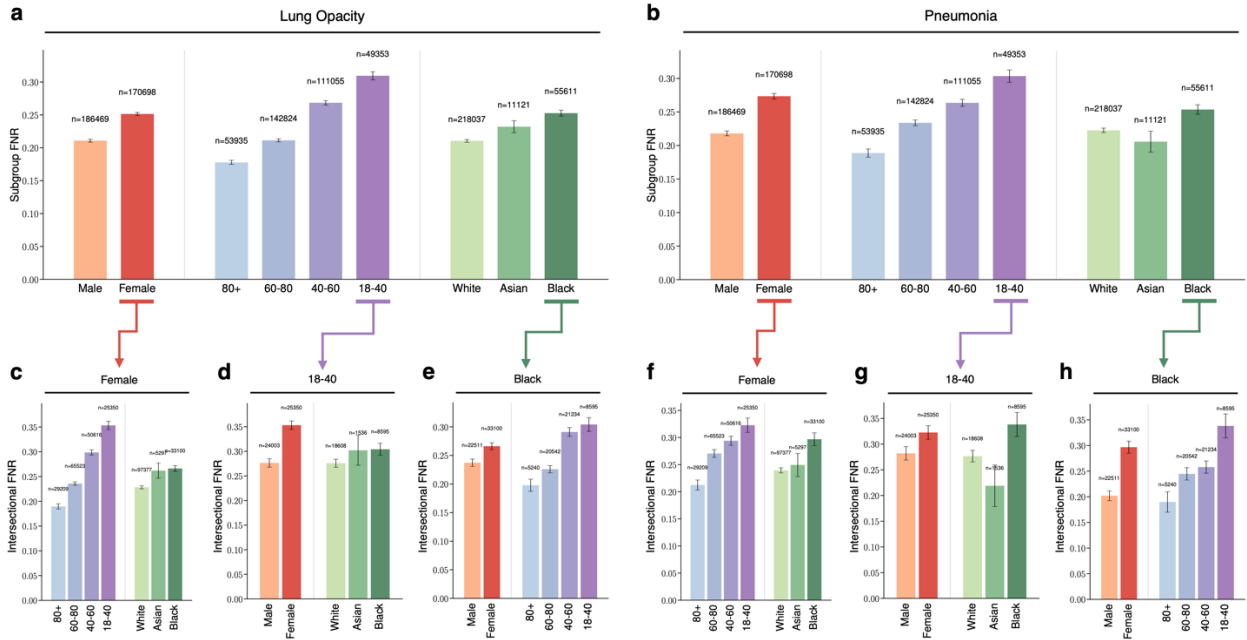

**Extended Data Figure 3. Underdiagnosis disparities on different pathologies across subgroups of sex, age, race, and intersectional groups in the MIMIC dataset. a,** The underdiagnosis rate for "Lung Opacity" in the indicated patient subpopulations. **b,** The underdiagnosis rate for "Pneumonia" in the indicated patient subpopulations. **c, d, e,** Intersectional underdiagnosis rates for "Lung Opacity" in female patients (**c**), patients aged 18–40 years (**d**), and Black patients (**e**). **f, g, h,** Intersectional underdiagnosis rates for "Pneumonia" in female patients (**f**), patients aged 18–40 years (**g**), and Black patients (**h**). Error bars indicate 95% confidence intervals estimated using non-parametric bootstrap sampling (n=1,000) are shown.

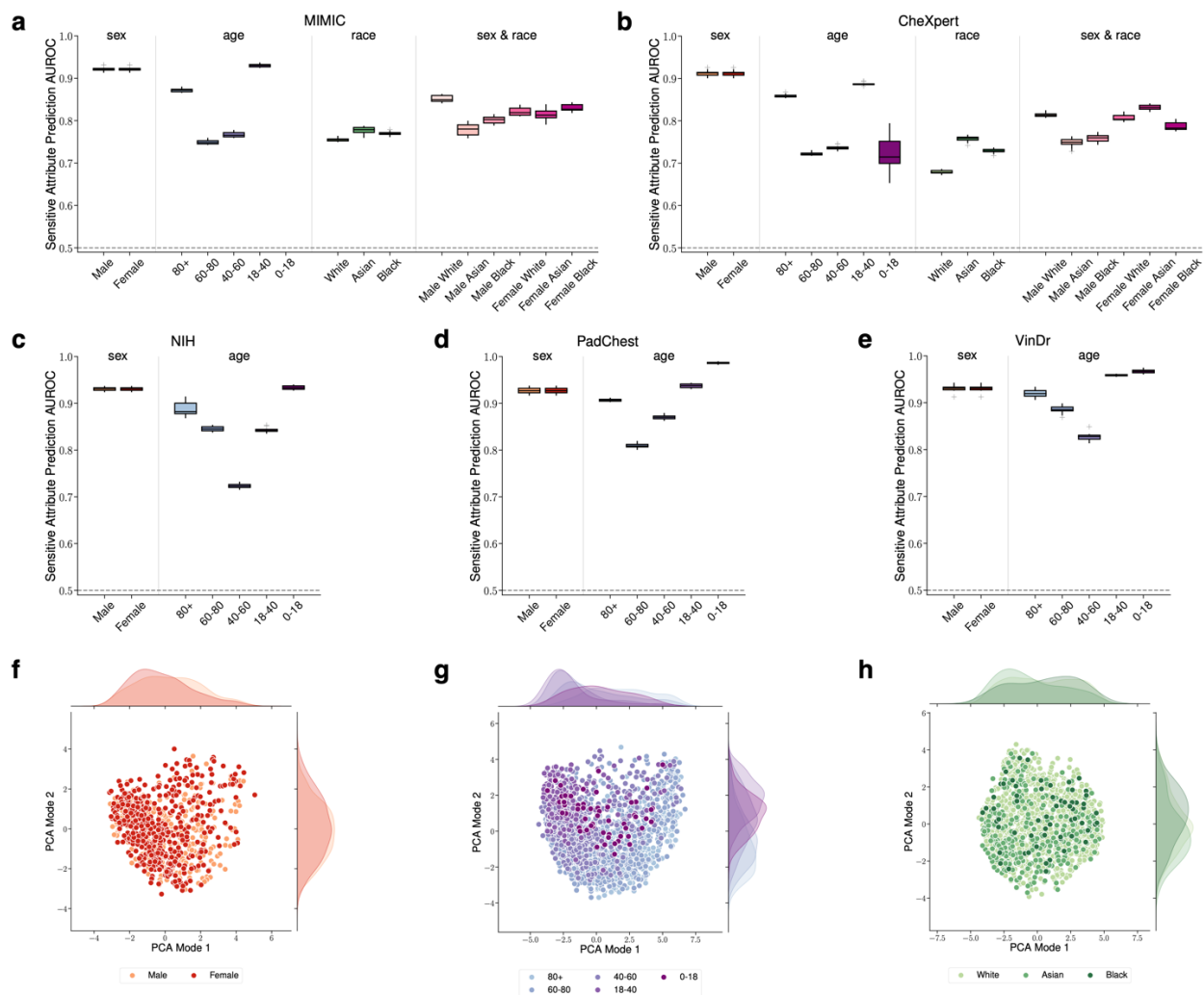

**Extended Data Figure 4. Algorithmic encoding of sensitive attributes in the foundation model. a, b, c, d, e,** Prediction AUROC of different sensitive attributes including age, sex, race, and intersectional groups, across five datasets including MIMIC (**a**), CheXpert (**b**), NIH (**c**), PadChest (**d**), and VinDr (**e**). We train a linear attribute prediction head using logistic regression on top of the penultimate layer of the model, with the model weights frozen. **f, g, h,** PCA visualization of the learned features in the penultimate layer of the model. We visualize the feature distribution on the randomly subsampled CheXpert dataset (n=2,000) for different attributes including sex (**f**), age (**g**), and race (**h**).

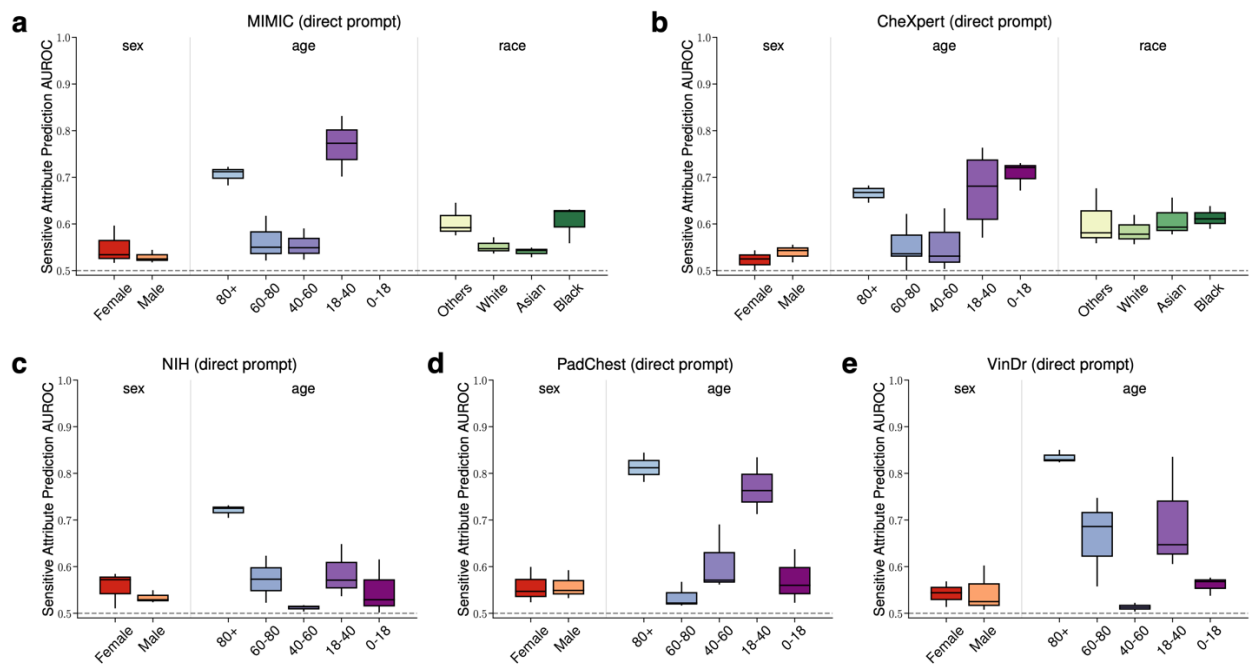

**Extended Data Figure 5. Direct Attribute prediction AUROC of the foundation model across different datasets. a, b, c, d, e,** We utilize textual prompts encompassing demographic information (e.g., "The patient's gender is male.") to assess the attribute prediction accuracy on the MIMIC (**a**), CheXpert (**b**), NIH (**c**), PadChest (**d**), and VinDr (**e**) datasets. Error bars indicate 95% confidence intervals estimated using non-parametric bootstrap sampling (n=1,000) are shown.

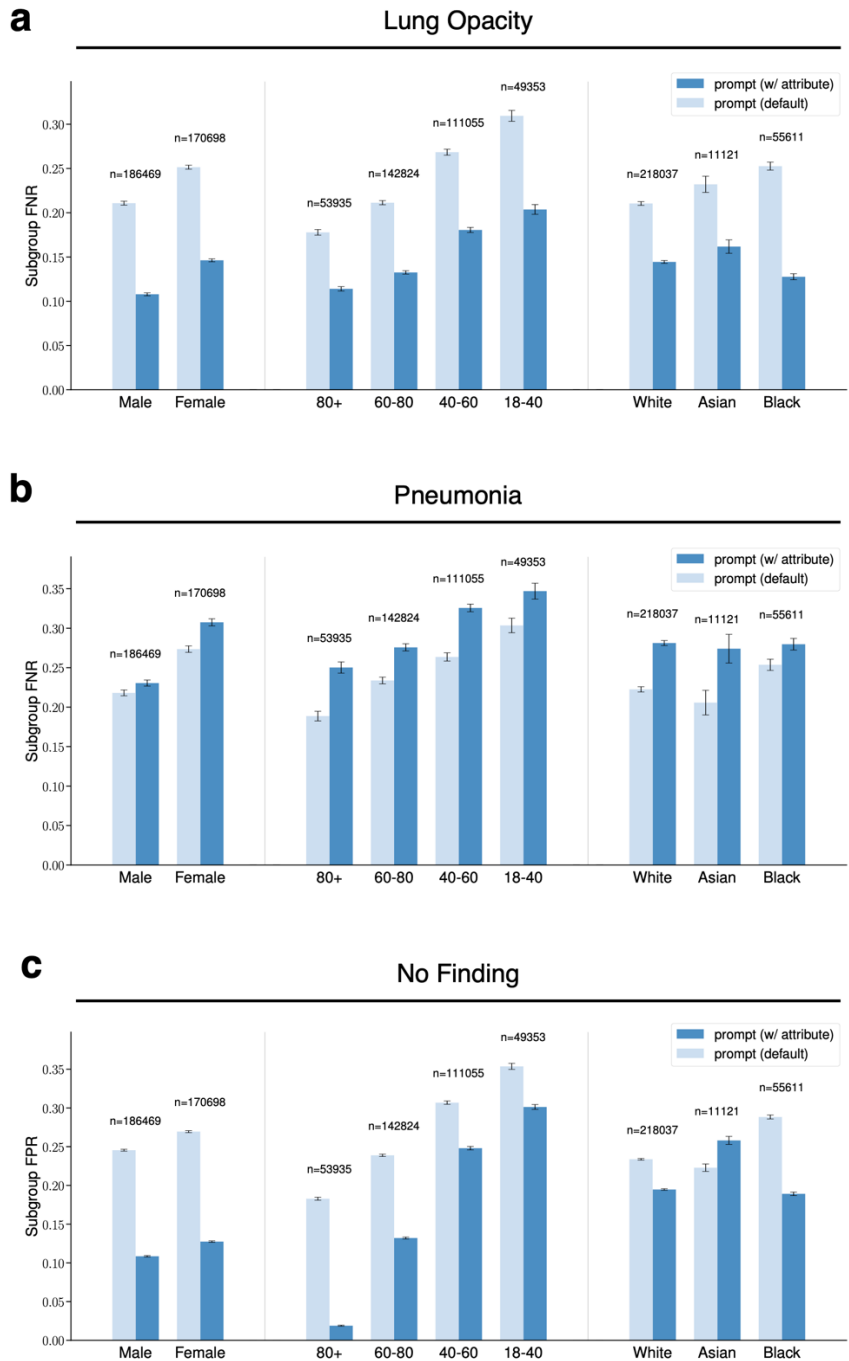

**Extended Data Figure 6. Model fairness intervention by incorporating demographic details into the input prompt. a, b, c,** Performance across subgroups before and after introducing the sensitive demographic details into the prompt, for "Lung Opacity" (a), "Pneumonia" (b), and "No Finding" (c). We proposed to intervene the model prediction over subgroups by including demographic information in the input texts (e.g., "Does this female patient have Pneumonia?"). After this intervention, the model displays reduced demographic biases for certain conditions like "Lung Opacity", but not for others like "Pneumonia". Error bars indicate 95% confidence intervals estimated using non-parametric bootstrap sampling (n=1,000) are shown.